# Quantum resonant tunnelling enhances hydrogen bond rotation of confined water

Le Jin[a], Xinrui Yang[a], Yu Zhu[a], Zhiyuan Zhang[a], Rui Liu[a] and Zhigang Wang[a,1]

[a]Institute of Atomic and Molecular Physics, Jilin University, Changchun 130012, China;

[1]Correspondence authors: wangzg@jlu.edu.cn (Z. W.).

## Abstract

Many studies have revealed that confined water chain flipping is closely related to the spatial size and even quantum effects of the confinement environment. Here, we show that these are not the only factors that affect the flipping process of a confined water chain. First-principles calculations and analyses confirm that quantum tunnelling effects from the water chain itself, especially resonant tunnelling, enhance the hydrogen bond rotation process. Importantly, resonant tunnelling can result in tunnelling rotation of hydrogen bonds with a probability close to 1 with only 0.597 eV provided energy. Compared to sequential tunnelling, resonant tunnelling dominants water chain flipping at temperatures up to 20 K higher. Additionally, the ratio of the resonant tunnelling probability to the thermal disturbance probability at 200 K is at least ten times larger than that of sequential tunnelling, which further illustrates the enhancement of hydrogen bond rotation brought about by resonant tunnelling.

## Introduction

To achieve effective regulation of the transport process of confined water, an important goal is to deeply understand the hydrogen bond (H-bond) rotation mechanism at the atomic level. Quantum effects have been reported to be essential and should be carefully considered on a microscopic scale,[1-7] including quantum effects related to water, such as properties of electronic correlation,[8-9] nuclear quantum effects,[10-11] tunnelling,[12-13] and electron density delocalization.[14] This suggests that quantum effects play an important role in understanding the nature of water and even regulating its behaviour. More importantly, some studies show that the H-bond rotation of water is affected by quantum effects.[8, 15] As a microstructure, a hydrogen atom has a very small mass, which leads to the existence of tunnelling effects from the water chain itself during H-bond rotation. Moreover, a recent experiment revealed that quantum coherence can effectively improve the tunnelling effects.[1] This inspired us to explore the effect of the quantum properties of a water chain on the H-bond rotation process.

Currently, many studies have been performed on the transport of confined water,[16-19] including on the step-by-step rotation behaviour of confined water and its intrinsic mechanism.[20] Many studies have shown that the flipping of confined water is affected by the spatial size;[21-26] for example, the transport of confined water has a high conductivity,[24-25] and the water flux through a carbon nanotube (CNT) has a linear relationship with the radius of the CNT when the length is much larger than the radius.[26] However, due to the inherent complexity of the molecular system, the understanding of the mechanism of water chain flipping under confinement is still unclear. The experimental technique has been greatly improved compared with the past, which has resulted in water chains formed in CNTs with diameters as small as 0.548 nm, and an extremely high freezing temperature has been detected;[27-28] however, observing the details of the mechanism of water chain flipping in a short time, much less regulating the mechanism of water chain flipping, is still a great challenge. Considering that the tunnelling effects, especially the coherence, have a more fundamental significance in quantum physics, studying the resonance effects caused by the coherence that exists when a water chain flips in a confined space from the perspective of the basic principles of tunnelling is necessary.

In this paper, by means of quantum calculations, we study the possible flipping processes of a water chain under quasi-one-dimensional (1D) confinement and find that the quantum effects cannot be ignored. Significantly, quantum resonant tunnelling brings different effects than traditional tunnelling and general thermal processes. Here, the calculation results show that the H-bond rotation of the water chain in a 1D-CNT is obviously affected by tunnelling, especially resonant tunnelling. Compared with the tunnelling without considering quantum coherence, the tunnelling considering quantum coherence can achieve a higher probability with less provided energy, which increases the temperatures at which tunnelling dominants the water chain flipping by up 20 K. Therefore, our work opens a new perspective for the quantum regulation of water chain flipping in channels.

## Result and discussion

To study the possible flipping processes of a water chain under confinement, the water molecules in a CNT that form a 1D chain are examined (Figure. 1(a)). We give two reaction paths, and the corresponding potential energy surfaces (PESs) of flipping are shown in Figure. 1(b). There are five extreme points along each reaction path: two structures with lower energies (reactant and product, labelled $Min_1$ and $Min_2$), two transition state structures (labelled $TS_1$ and $TS_2$), and an intermediate structure (labelled Int). These structures are shown in Figure. 1(c). This suggests that the flipping of the water chain is a step-by-step process, and the formation of the intermediate is necessary, which is consistent with our previous study.[20]

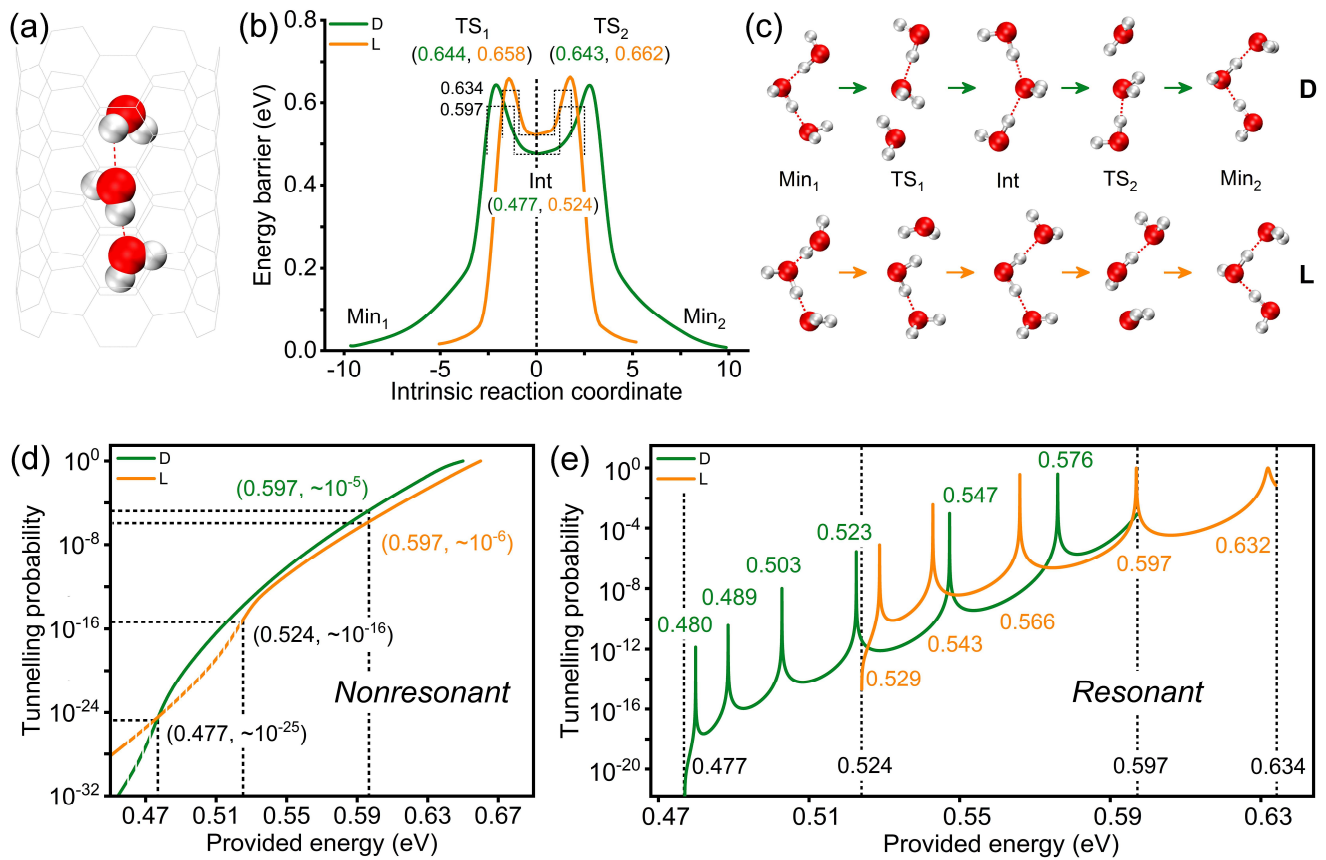

**Figure 1.** Different mechanisms by which the water chain in the CNT achieves reorientation. (a) Model of the confined water chain. (b) PESs of the water chain flipping. Dotted lines are the square energy barriers obtained by fitting the double barriers of the PESs. (c) Water chain structures corresponding to each extreme point on the PESs. (d) Variation in the quantum tunnelling probability with provided energy without considering coherence. Based on the Born-Oppenheimer approximation, the Wentzel-Kramers-Brillouin (WKB) approximation [29-30] is used for calculating the tunnelling probability for a single barrier using the following formula: $\boldsymbol{P = Exp\left[-\frac{2}{\hbar}\int_{x_1}^{x_2}\sqrt{2m(V(x)-E)}\,dx\right]}$. The tunnelling probability for double barriers is obtained by multiplying the tunnelling probability of two single barriers. The dotted and solid lines indicate the tunnelling probability at provided energies below and above the intermediate energies, respectively. (e) Variation in the quantum tunnelling probability with provided energy considering coherence. According to the square barriers, the energy

height provided are 0.477–0.597 eV and 0.524–0.634 eV, corresponding to the two reaction paths. The dotted lines represent the provided energy at the start or end of the two reaction paths. The tunnelling probability is defined as the ratio of the flow of particles out of the barrier to the flow of particles into the barrier as follows: $T = \frac{|C_{4N+1}|^2}{|C_1|^2}$. Green and yellow represent the reaction paths containing D-type and L-type defective intermediates, respectively.

For the two reaction paths, the main difference lies in the structure of the intermediate. In detail, when the water chain flips along the nanotube axis in the opposite direction to the H-bonds, the heights of the energy barriers of the two transition states to be overcome are approximately 0.660 eV, and an intermediate with an energy of approximately 0.524 eV is formed. This structure has the two hydrogen atoms in the middle water molecule forming H-bonds with the oxygen atoms of the two adjacent water molecules. We call this the intermediate with an L-type defect.[31] The H-bonds are oriented towards the middle water molecule. When the water chain flips along the direction of the H-bond, it needs to overcome two energy barriers with heights of approximately 0.640 eV. Additionally, the intermediate formed has an energy of approximately 0.477 eV.

Reorientation of the water chain can be achieved not only by step-by-step flipping but also by quantum tunnelling. Here, to obtain the sequential tunnelling probability for double barriers without considering the quantum coherence, the Wentzel-Kramers-Brillouin (WKB) approximation is applied [29-30]. We considered the difficulty of the water chain to achieve reorientation by quantum tunnelling at different provided energies. As shown in Figure. 1(d), with an increase in the provided energy, the tunnelling probability gradually increases. Taking the path containing the L-type defective intermediate as an example, the tunnelling probabilities are approximately $10^{-25}$, $10^{-16}$ and $10^{-6}$ when the provided energies are approximately 0.477 eV, 0.524 eV and 0.597 eV, respectively. For the two paths containing L-type and D-type defective intermediates, when the provided energy is approximately 0.477 eV, the two tunnelling probabilities are almost equal. However, when the provided energy is larger than 0.477 eV, the tunnelling probability of the water chain along the path containing the L-type defective intermediate is less than that along the path containing the D-type defective intermediate. For example, at the provided energy of 0.597 eV, the tunnelling probabilities are approximately $10^{-6}$ and $10^{-5}$ for the reaction paths containing the L-type and D-type defective intermediates, respectively. When the provided energy is less than 0.477 eV, the tunnelling probability of the water chain along the path containing the L-type defective intermediate is greater than that along the path containing the D-type defective intermediate. In addition to quantum tunnelling, thermal effects can also cause the water chain to flip. To clearly show the relationship between tunnelling and thermal disturbance with temperature and provided energy, we regard thermal disturbance and tunnelling as completely independent probability events. As shown in Figure. 2(a), we compare the probabilities of the water chain achieving reorientation by quantum tunnelling and thermal disturbance (25–200 K) when the provided energies at the double barriers of the two reaction paths are given. With an increase in the temperature, the probability of water chain flipping by thermal disturbance increases. When the temperature is lower than 84 K, quantum tunnelling is more dominant than thermal disturbance. At temperatures above 120 K, thermal disturbance plays a dominant role, rather than quantum tunnelling. Therefore, we infer that at room temperature, thermal disturbance is more likely to cause the water chain to flip.

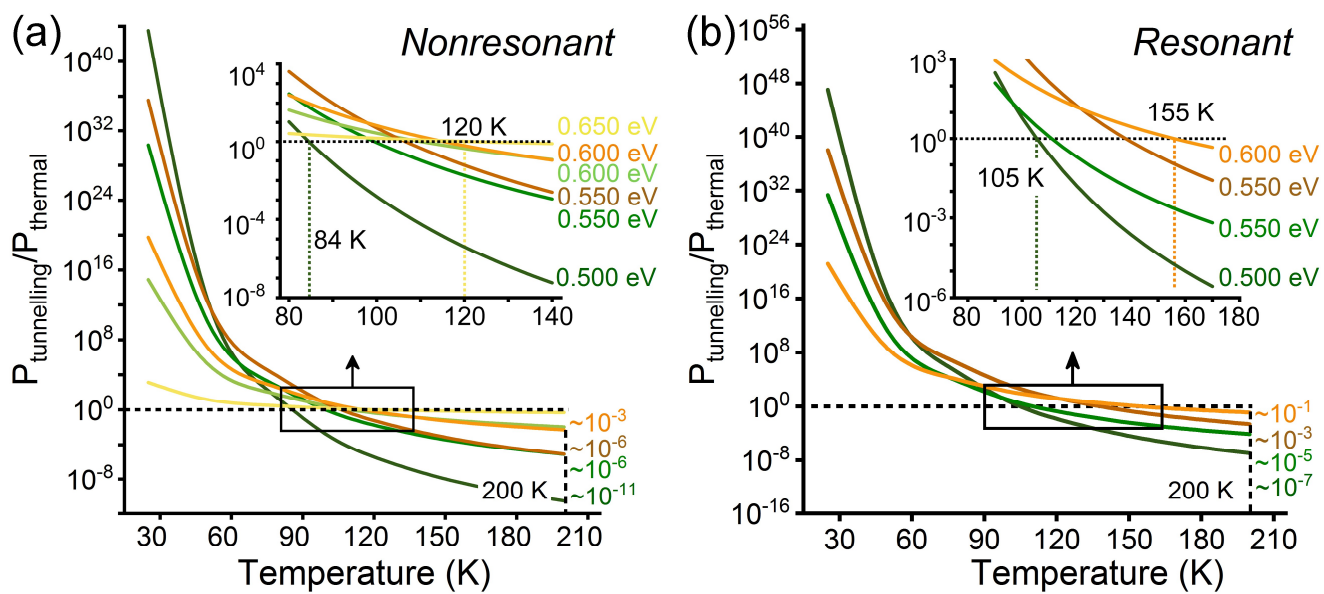

**Figure 2.** Ratio of the tunnelling probability to the thermal disturbance probability for the water chain in the CNT to achieve reorientation at certain provided energies and different temperatures. (a) Coherence is not considered in the tunnelling probability calculation. (b) Coherence is considered in the tunnelling probability calculation. The temperature range is 25–200 K, and the interval is 25 K. "~" is the magnitude.

Remarkably, due to the narrow width (approximately 1.9 Å and 2.5 Å for the reaction paths containing the L-type defect and D-type defect, respectively) of the well between the double barriers along the reaction path, the quantum coherence should not be ignored. Therefore, a discussion of the quantum tunnelling effects while considering coherence, that is, resonant tunnelling, is necessary. The tunnelling probability is defined as the ratio of the flow of particles out of the barrier to the flow of particles into the barrier. As shown in Figure. 1(e), with an increase in the provided energy, the tunnelling probability increases and decreases at times, showing a nonlinear change. This is significantly different from the quantum tunnelling probability without considering coherence. In addition, according to the peaks in Figure. 1(e), compared with quantum tunnelling without coherence, quantum tunnelling with coherence can achieve a higher probability at smaller provided energy. Taking the path containing the L-type defective intermediate as an example, the probabilities of quantum tunnelling without and with consideration of coherence are approximately $10^{-15}$, $10^{-12}$, $10^{-10}$, $10^{-6}$, and $10^{-3}$ and $10^{-6}$, $10^{-3}$, $10^{-1}$, $10^{0}$, and $10^{0}$ when the provided energies at the double barrier are approximately 0.529 eV, 0.543 eV, 0.566 eV, 0.597 eV and 0.632 eV, respectively. Furthermore, we compare the probabilities of the water chain achieving reorientation via two different mechanisms (quantum tunnelling considering coherence and thermal disturbance), as shown in Figure. 2(b). The results show that when the temperature is lower than 105 K, tunnelling is dominant compared with thermal disturbance. When the temperature is higher than 155 K, thermal disturbance gradually becomes dominant. This temperature is approximately 20 K higher than that without considering coherence, indicating that the neglect of quantum coherence could lead to underestimation of the tunnelling capability. In addition, the ratios of the tunnelling probability to the thermal disturbance probability at 200 K are analysed (as shown in Figure. 2). For example, the ratios with coherence and without coherence are approximately $10^{-11}$ and $10^{-7}$ when the provided energy is 0.500 eV for the reaction path containing the D defect, respectively. This further illustrates that resonant tunnelling is more favourable to H-bond rotation of the confined water chain than sequential tunnelling. Importantly, resonant tunnelling has been experimentally observed in hydrogen atom systems at energies of $10^{-3}$ eV in previous work [1]. Our work is expected to provide theoretical guidance for the quantum regulation of water chain flipping in channels in future experiments when the accuracy of the provided energy is approximately $10^{-3}$.

## Conclusion

In this paper, we propose for the first time the quantum tunnelling effects from the water chain itself under ideal conditions, which enhance the H-bond rotation process. Importantly, compared with the tunnelling without considering coherence, the tunnelling considering coherence can achieve a higher probability given less provided energy. The study of tunnelling properties enriches the understanding of quantum coherence for H-bonds and hopefully will enable quantum regulation of the H-bond rotation mechanisms for confined water to be achieved. In the future, we will also further consider issues such as vibrational decoherence.

## Methods

For quantum calculations, the empirical-dispersion-corrected hybrid Perdew-Burke-Ernzerhof (PBE0-D3) method of density functional theory was carried out in the Gaussian 09 package [32-34]. The basis sets 6-311+G (d, p) and 6-31G (d) were used for water and the CNT, respectively. The armchair-type single-walled (6, 6) CNT was employed, and the diameter and length were set to 8.20 Å and 20 Å, respectively. After CNT preoptimization, we froze all of the atoms of the CNT to ensure constant confinement effects on the water. In the CNT, the water chain was along the tube axis, and the molecules were connected to each other by H-bonds. Based on the different initial geometries of the water chain, we searched the structures for extreme points (including equilibrium and transition states) in the rotation process of water molecules in the CNT and traced the reaction paths of the flipping process for the water chain according to the intrinsic reaction coordinate [35-36]. The reduced masses are 1.0834 amu and 1.0955 amu for the water chain flipping along the path containing L-type and D-type defective intermediates.

For the formula without considering quantum coherence, $\hbar$ is the reduced Planck constant, $V(x)$ represents the potential energy surface PES function with the coordinate $x$ as the variable, expressed by a Gaussian fitting function, and $x_1$ and $x_2$ are the two coordinates when $V(x)$ and $E$ are equal. For the formula considering quantum coherence, the steady-state Schrödinger equation of $N$ multiple barriers is strictly solved. For the convenience of calculation, the equivalent square barrier is used to fit the PES. In detail, we first take the half-height width of the left-side barrier as the square barrier width to obtain the approximate equivalent square barrier. Then, the local minimum value is taken as the axis of symmetry to obtain the ideal double barrier model. This equivalent square barrier method is an approximation based on the principle that original barrier and approximate square barrier have the same tunnelling probability. The general expression of the Schrödinger equation can be written as follows:

$$\psi_j = C_{2j-1} \exp(ik_j x) + C_{2j} \exp(-ik_j x)$$

in which $C_{2j-1}$ represents the transmission amplitude, $C_{2j}$ is the reflection amplitude and $k_j$ is the wavenumber.

For the thermodynamic model, the Boltzmann distribution is used to describe the probability of crossing the barrier from the classical perspective as follows:

$$P_{thermal} = Exp\left(-\frac{\Delta E}{kT}\right)$$

where $k$ is the Boltzmann constant and $T$ is the temperature. $\Delta E$ represents the relative energy between the initial provided energy and the barrier peak. Additionally, to compare the classical and quantum methods of crossing the barrier, the ratio of the quantum tunnelling probability to the thermal disturbance probability is studied.

## Notes

The authors declare no competing financial interest.

## Author Contributions

## Acknowledgment

This work was supported by the National Natural Science Foundation of China (under grant number 11974136 and 11674123). Z. W. also acknowledges the High-Performance Computing Center of Jilin University and National Supercomputing Center in Shanghai.